\documentclass[twocolumn,english,aps,prl,superscriptaddress,floats,nobibnotes]{revtex4-1}
\usepackage[latin9]{inputenc}
\usepackage{color}
\usepackage{babel}
\usepackage{bm}
\usepackage{bbm}
\usepackage{amsmath}
\usepackage{amssymb}
\usepackage{graphicx}
\PassOptionsToPackage{version=3}{mhchem}
\usepackage{mhchem}
\usepackage{braket}
\usepackage[unicode=true,
 bookmarks=false,
 breaklinks=false,pdfborder={0 0 1},backref=false,colorlinks=false]
 {hyperref}
\hypersetup{
 colorlinks,linkcolor=blue,citecolor=blue,urlcolor=blue}

\makeatletter
\usepackage{babel}
\PassOptionsToPackage{version=3}{mhchem}

\usepackage{newfloat}
\usepackage{latexsym}
\usepackage{dcolumn}
\usepackage{float}

\usepackage{bm}
\usepackage{babel}

\makeatother

\begin{document}

\title{Superconductivity from collective excitations in magic angle twisted bilayer graphene}

\author{Girish Sharma}
\affiliation{Centre for Advanced 2D Materials, National
University of Singapore, 6 Science Drive 2, 117546, Singapore}
\affiliation{Department of Physics, National
University of Singapore, 2 Science Drive 3, 117551, Singapore}

\author{Maxim Trushin}
\affiliation{Centre for Advanced 2D Materials, National
University of Singapore, 6 Science Drive 2, 117546, Singapore}

\author{Oleg P. Sushkov}
\affiliation{School of Physics, The University of New South Wales, Sydney 2052, Australia}

\author{Giovanni Vignale}
\affiliation{Centre for Advanced 2D Materials, National
University of Singapore, 6 Science Drive 2, 117546, Singapore}
\affiliation{Yale-NUS College, 16 College Avenue West, 138527, Singapore}
\affiliation{Department of Physics and Astronomy, University of Missouri, Columbia, Missouri 65211, USA}

\author{Shaffique Adam}
\affiliation{Centre for Advanced 2D Materials, National
University of Singapore, 6 Science Drive 2, 117546, Singapore}
\affiliation{Department of Physics, National
University of Singapore, 2 Science Drive 3, 117551, Singapore}
\affiliation{Yale-NUS College, 16 College Avenue West, 138527, Singapore}

\date{\today}
%%TC:ignore
\begin{abstract}
A purely electronic mechanism is proposed for the unconventional superconductivity recently observed in twisted bilayer graphene (tBG) close to the magic angle. 
Using the Migdal-Eliashberg framework on a one parameter effective lattice model for tBG we show that a superconducting state can be achieved by means of collective electronic modes in tBG.   We posit robust features of the theory, including an asymmetrical superconducting dome and the magnitude of the critical temperature that are in agreement with experiments.
\end{abstract}
%%TC:endignore

\maketitle
\textit{Introduction:} The remarkable experimental observations of superconducting and insulating phases in twisted bilayer graphene (tBG)~\cite{cao_unconventional_2018, cao_correlated_2018, yankowitz_2019_tuning} close to the magic angle ushers a new paradigm in attempts to study strongly correlated phases of matter by bandstructure engineering. On stacking two graphene monolayers at a small relative twist angle $\theta$, emerges a large wavelength moir\'e superlattice potential.  Further, band calculations reveal~\cite{bistritzer2011moire} that there is substantial renormalization of the Fermi velocity giving rise to flat bands around half filling
once $\theta$ is sufficiently small. At some very specific `magic angles' ($\theta_M\sim 1.1^{\circ}$), the Fermi velocity is predicted to vanish facilitating strong electronic correlations. 
It is therefore plausible that the superconducting instability observed in twisted bilayer graphene is a consequence of these correlations.

\begin{figure}
    \centering
    \includegraphics[scale = 0.5]{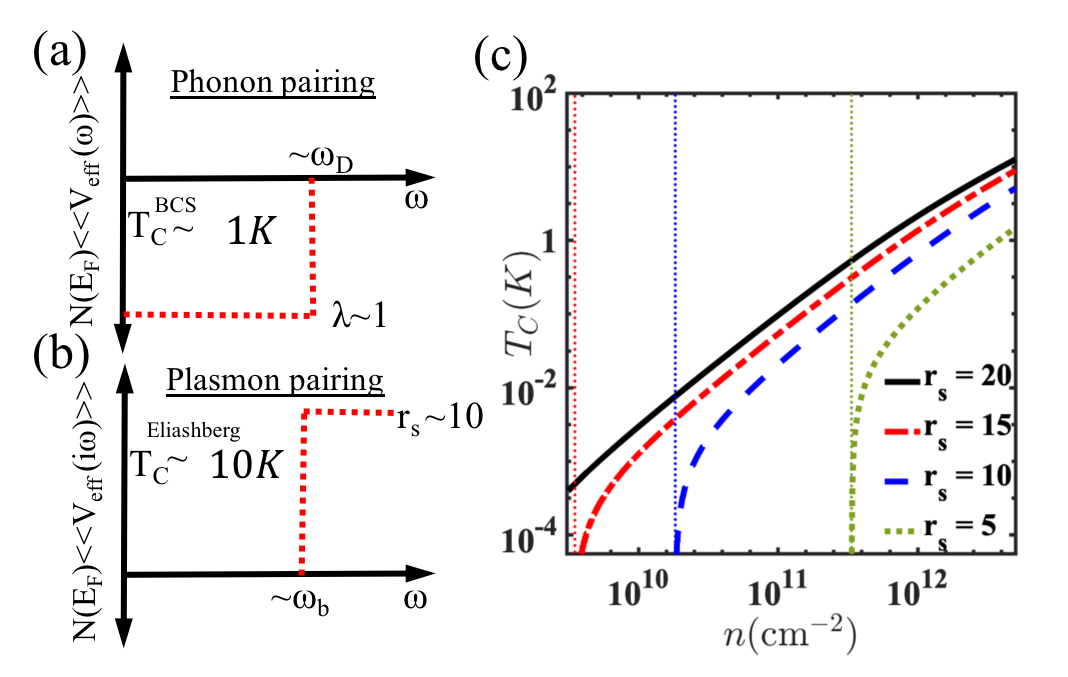}
\caption{The single-mode model for the dimensionless pairing potential $N(E_F)\langle\langle V_{\text{eff}}(\omega)\rangle\rangle$ ($N(E_F)$ being the density of states and $\langle\langle V_{\text{eff}}(\omega)\rangle\rangle$ the effective interaction) in tBG for (a) moir\'{e} phonons (with Debye frequency $\omega_D$) within BCS theory~\cite{lian2019twisted}, and (b) collective electronic excitation (with frequency $\omega_b$) such as a plasmon within Eliashberg theory. In this system with purely repulsive interactions, the frequency dependence of the pairing potential can induce superconductivity. 
%The plasmon coupling can be at least ten times stronger, and yield a critical temperature to be $\sim 10K$. (This contradicts to fig.1(c) and fig. 3 within reasonable range of n. --- M.T.)
(c) $T_C$ evaluated within the single-mode approximation (Eq.~\ref{TC}) for a Dirac model with renormalized $v_F\sim 1.5\times 10^4$ms$^{-1}$, $\omega_b \sim 15$ meV. Below a threshold density (indicated by the dotted lines), no solution for $T_C$ exists. }
    \label{fig1_schematic}
\end{figure}

Given the rich underlying physics at play, a lot of theoretical effort has already been devoted towards understanding the Mott-like physics of strongly correlated electrons in tBG~\cite{xu_topological_2018,roy_unconventional_2018,po_origin_2018,koshino_maximally_2018,kang_symmetry_2018,padhi_doped_2018,guo_pairing_2018,liu_chiral_2018,isobe_unconventional_2018,you_superconductivity_2018,gonzalez_kohn-luttinger-2019,xie_on_2018, laksono2018singlet}, and attributing the observed superconducting state to weak electron-phonon coupling~\cite{lian2019twisted, choi_strong_2018, wu_theory_2018, yudhistira2019gauge, wu2018phonon}. It is however not obvious that the latter is a conventional Bardeen-Cooper-Schriefer (BCS) state even though the phonon mediated superconducting pairing is amplified in tBG due to the enhanced density of states at the Fermi surface (see Fig.~\ref{fig1_schematic}). Neither is there a universal consensus or understanding that Mott-like physics is in action. 
In contrast, Coulomb interactions and associated plasmonic effects are known to be
very strong in graphene~\cite{grigorenko2012graphene}, and theoretically
~\cite{grabowski1984superconductivity}, may lead to a superconducting state.
In this Letter we explore the possibility of unconventional superconductivity in tBG mediated by the purely collective electronic modes. 
%It is a fair statement that the nature of the insulating phase, normal phase, the symmetry and mechanism of the superconducting pairing are all largely open unsettled questions. 

We first emphasize that the electronic mechanism is quite different from the standard BCS interaction, as no phonon modes are necessary. The collective oscillations can generate an effective dynamic attractive interaction, thus Cooper-pairing two electrons. Starting from a one-parameter effective lattice model for tBG \cite{lewandowski2019intrinsically} we first calculate the dynamical polarization function $\Pi(\mathbf{q},i\omega)$ and thereby the dynamically screened Coulomb interaction $V(\mathbf{q},i\omega)$. 
Following Grabowski and Sham~\cite{grabowski1984superconductivity}, we average the interaction kernel to a dimensionless momentum-independent interaction parameter, $\lambda_{nm}$,
and obtain the gap equation as follows:
\begin{align}
    \Delta(i\omega_n) = -\frac{2T_C}{E_F} \sum\limits_m \frac{ \arctan\left[E_F/(Z_m \omega_m)\right]}{(Z_m \omega_m)/E_F} \lambda_{nm} \Delta(i\omega_m),
    \label{Eq:gap4}
\end{align}
where $\omega_n = (2n+1)\pi T_C$ is the Matsubara frequency, and $Z_m$
is a renormalizing function calculated below.
Solving Eq. (\ref{Eq:gap4}) we find the superconducting critical temperature $T_C$ for various twist angles $\theta$ and carrier concentrations linked to the Fermi energy $E_F$ via density of states.
The prominent predictions of our theory are: (i) the plasmon-assisted Cooper pairing
is much stronger than the conventional phonon-related one in tBG,
(ii) the superconducting state does not occur at low electron concentrations
but is prominent at electron densities around $n=10^{12}$ cm$^{-2}$, 
(iii) the obtained critical temperature is of $\mathcal{O}(10 K)$, similar to that observed in experiments~\cite{cao_unconventional_2018, cao_correlated_2018,yankowitz_2019_tuning} and larger than those calculated by the phonon contribution~\cite{lian2019twisted}, (iv) our theory is not limited by particular plasmon modes as our model generically captures the effect of plasmons (whenever well-defined), but also the other density-fluctuation excitations.
Our work specifically points out to the mechanism for superconductivity that is likely to be at play in tBG. 
%Vertex corrections are ignored because they are likely to be insignificant when $\alpha = e^2/\kappa \hbar v_F$ is large~\cite{takada1992insignificance}.

\textit{Phonons or plasmons?} The striking scenario which develops at small twist angles is that the relevant electronic energy scales are shrunk to the order of a few meV  (similar to the acoustic phonon energy scale), while the density of states is amplified. This raises the intriguing possibility of observing collective electronic modes in the same energy window. For example, it was recently~\cite{lewandowski2019intrinsically} pointed out that plasmon modes in tBG can remain intrinsically undamped protected by the bandgap between the flat bands and higher bands. It is also worth noting the possibility to have hybrid acoustic phonon and plasmon modes still uncommon in condensed matter systems.

Let us compare the phonon and plasmon mediated superconductivity mechanisms
in tBG with $\theta$ close to $\theta_M$. 
Here, we assume a simplest two-dimensional massless Dirac fermion model
with the renormalized Fermi velocity $v_F$.
The Fermi-surface is then just a circle of radius $k_F$, and
the averaged electron-phonon coupling constant is
$\lambda = \zeta^2\pi k_F /(\hbar v_F \rho c_{ph}^2)\sim\mathcal{O}(1)$, where $\zeta$, $\rho$, and $c_{ph}$ are deformation potential, mass density and the sound velocity, respectively.
To the first approximation we can assume the attractive pairing potential to be finite only below the Debye frequency $\omega_D$ (see Fig.~\ref{fig1_schematic}a).
The McMillan formula then suggests $T_C\sim \omega_D \exp(-1/\lambda)\sim \mathcal{O}(1K)$, as also shown recently~\cite{lian2019twisted}. In contrast,
the plasmon-mediated mechanism suggests that the dynamic Coulomb interaction $V(\mathbf{q},i\omega)$ is responsible for the superconducting pairing.
For such pairing the dynamical nature of the coupling is crucial
because the Coulomb interaction is screened by the dielectric function $\epsilon(q,i\omega) = 1+e^2 E_F q/(2\kappa\omega^2)$~\cite{wunsch2006dynamical, hwang_dielectric_2007} 
providing stronger interactions in the high-frequency limit.
Here, the renormalized dielectric constant $\kappa\sim 12$ 
accounts for effects of interband polarization in tBG \cite{lewandowski2019intrinsically}. 
The corresponding dimensionless coupling $\lambda_{nm}$
reduces to the bare Coulomb interaction characterized by $r_s = e^2 / (\kappa \hbar v_F)$ in the high frequency limit, while remaining zero below a certain 
energy determined by $\omega_b$ \cite{grabowski1984superconductivity, canright1989superconductivity}.
This behavior being quite opposite to the phonon-assisted pairing 
is schematically shown in Fig.~\ref{fig1_schematic}b.

Equation (\ref{Eq:gap4}) is still too complicated, as it in fact represents an infinite number of
coupled equations. We design a simple analytically tractable model by 
reducing the number of equations to three considering only the terms with $m=0, \pm M$, where $M\gg 1$. Neglecting the self-energy corrections for now (these will be included later on) and setting the diagonal elements $\lambda_{mm}=0$ we arrive at the following equation for $T_C$:
\begin{equation}
 r_s\frac{2T_C}{\omega_M}\arctan\left(\frac{E_F}{\omega_M}\right)
 \left[\frac{4r_s}{\pi}\arctan\left(\frac{E_F}{\pi T_C}\right) - 1\right] =1.
 \label{TC}
\end{equation}
We use the single-mode Eq.~\ref{TC} with $M\sim \omega_b/E_F$ to estimate $T_C$ in Fig. \ref{fig1_schematic}c.
In the low-$T_C$ limit ($T_C \ll E_F$) and strong coupling ($r_s\gg 1$), we find
an elegant expression
\begin{equation}
\arctan\left(\frac{E_F^2}{2\pi\omega_b T_C}\right)=\frac{\pi}{2}\frac{\omega_b}{r_s^2 E_F},
\label{main}
\end{equation}
that can be seen as a plasmonic analogue of the McMillan formula for $T_C$.
Since the left-hand side of Eq. (\ref{main}) cannot exceed $\pi/2$
the solution for $T_C$ exists if and only if $\omega_b/(r_s^2 E_F)<1$,
i.e. the electron concentration and coupling strength values must be large enough.
This is indeed the case in tBG.
Estimating $\omega_b\sim 15$ meV, $v_F\sim 1.5\times10^4$ m/s, we find  $r_s\sim 12$, $E_F=\hbar v_F \sqrt{\pi n/2}\sim 1.5$ meV
(for electron density $n\sim  1.5\times 10^{12}$ cm$^{-2}$), and $M\sim 10$, resulting in a reasonable value $T_C\sim 2.6$ K matching the observations \cite{cao_unconventional_2018}. 
In conventional graphene $r_s \sim 1$, $\omega_b > E_F$ \cite{hwang_dielectric_2007},
and solution of Eq. (\ref{main}) never exists making the plasmon model
tBG-specific. This is consistent with the fact that superconductivity has been
observed in tBG but not in monolayer graphene.

%The fact that using comparable approximations the plasmonic $T_C$ is an order of magnitude larger than the phononic one strongly suggests that plasmon-assisted pairing, which we investigate in detail below, is relevant for the observed experimental superconductivity.  

\textit{Dynamical Coulomb interaction:} To go beyond the single-mode model, we consider the following  one-parameter effective nearest-neighbor tight binding Hamiltonian defined on a hexagonal lattice, which mimics the low energy Hamiltonian of twisted bilayer graphene~\cite{lewandowski2019intrinsically}
\begin{align}
    H &= \sum\limits_{\mathbf{k}}{h_\mathbf{k} c^\dagger_{\mathbf{k},2}c_{\mathbf{k},1} + h^*_{\mathbf{k}}c^\dagger_{\mathbf{k},1}c_{\mathbf{k},2}},\nonumber\\
    h_{\mathbf{k}} &= t_{\hbox{eff}} \sum\limits_{\delta_\mathbf{j}} e^{i \mathbf{k}\cdot \delta_\mathbf{j}},
\end{align}
where $c_{\mathbf{k},\eta}$ and $c^\dagger_{\mathbf{k},\eta}$ are the annihilation and creation operators for electrons with momentum $\mathbf{k}$ in the Brillouin zone (BZ) on the  sublattice $\eta$. The effective hopping matrix element $t_{\hbox{eff}} = W/3$ corresponds to the bandwidth ($W$) of the nearly flat bands in tBG close to the magic angle. The summation $\delta_{\mathbf{j}}$ is over the nearest neighbors $\delta_\mathbf{j}=(\cos(2\pi j/3), \sin(2\pi j/3)) \tilde{a}\sqrt{3}$, where $\tilde{a} = (a/2 \sin(\theta/2))$ is the periodicity of the moir\'e superlattice. The constants $a=2.46$\AA$ $ is the graphene lattice constant, while $\theta$ is the twist angle. For our calculations we obtain the bandwidth $W$ from the tBG continuum model~\cite{bistritzer2011moire}. The advantage of the above tight-binding model is that it reproduces the symmetry of the actual tBG and has a natural ultraviolet smooth cutoff scale $W$. The divergent density of states at the van Hove singularity is also manifested in this model, which will be important for our analysis.   %The bandwidth $W$ is an important parameter for several reasons. For instance, plasmons can stay intrinsically undamped above $\omega = 2W$~\cite{lewandowski2019intrinsically}. However specific consideration of plasmon damping is not a matter of concern for us because our interaction is global, i.e., it not only includes acoustic plasmons (whenever well-defined), but also the other density-fluctuation excitations (for example electron-hole pairs). 
The eigenvalues are $E^n_\mathbf{k} = n |h_\mathbf{k}|$ and the four-fold degenerate eigenstates are $\psi^\dagger_{n,\mathbf{k}} = (2^{-\frac{1}{2}})[e^{-in\arg (h_\mathbf{k})}, 1]$, where $n=\pm 1$ is the band index. 

We are particularly interested in the dynamic polarization function $\Pi(\mathbf{q},i\omega)$, which is given by 
\begin{align}
    \Pi(\mathbf{q},i\omega) = 4 \sum\limits_{\mathbf{k}}\sum\limits_{m,n}\frac{(f^n_{\mathbf{k}+\mathbf{q}} - f^m_{\mathbf{k}}) F^{nm}_{\mathbf{k},\mathbf{k}+\mathbf{q}}}{E^n_{{\mathbf{k} +\mathbf{q}}} - E^m_\mathbf{k} - i\omega }.
\end{align}
where $n$, $m$ are the band indices, $f^n_\mathbf{k}$ is the Fermi-Dirac distribution, and $F^{nm}_{\mathbf{k},\mathbf{k}+\mathbf{q}}   =  |\psi^\dagger_{n,\mathbf{k}+\mathbf{q}} \psi_{m,\mathbf{k}} |^2$ is the graphene chirality factor. Since $h_\mathbf{k}$ has a  complicated momentum dependence, we  evaluate the above function  numerically expanding up to second order in $\mathbf{q}$. As recently shown~\cite{lewandowski2019intrinsically}, this is actually sufficient to describe collective excitations such as plasmons over a wide range of energy scales $0<\omega<2 W$.  The dynamical polarizability is the used to evaluate the dielectric constant within the random phase approximation (RPA) as $\epsilon(\mathbf{q},i\omega) = 1 - V_\mathbf{q} \Pi(\mathbf{q},i\omega)$, where $V_\mathbf{q} = 2\pi e^2/\kappa q$ is the bare Coulomb interaction. Finally, the dynamically screened Coulomb interaction is given by $V(\mathbf{q},i\omega) = V(\mathbf{q}) / \epsilon(\mathbf{q},i\omega)$. The dynamical dielectric constant $\epsilon(\mathbf{q},i\omega)$ is related to the collective propagator $\chi^{co}(\mathbf{q},i\omega)$  as $\epsilon(\mathbf{q},i\omega)^{-1} = 1+V(\mathbf{q})\chi^{co}(\mathbf{q},i\omega)$.

\textit{The gap equation:} 
The Migdal-Eliashberg superconductivity theory suggests the following gap equation~\cite{eliashberg1960interactions}
\begin{align}
\Delta(\mathbf{p},i\omega_n) = -T_C \sum_{\mathbf{k},m}[ V(\mathbf{p}&-\mathbf{k},i\omega_n-\omega_m) B (\mathbf{k},i\omega_m)\nonumber\\
&\Delta(\mathbf{k},i\omega_m)],
\label{Eq_gap1}
\end{align}
where the product of the Green's functions $B(\mathbf{k},i\omega_m)$ given by
\begin{align}
B(\mathbf{k},i\omega_m) &= G(\mathbf{k},i\omega_m) G(-\mathbf{k},-i\omega_m), \nonumber\\
    G(\mathbf{k},i\omega_m) &= \frac{1}{i\omega_m - \epsilon_k - \Sigma(\mathbf{k},i\omega_m)}. 
    \label{Eq_gap2}
\end{align}
Here, $\Sigma(\mathbf{k},i\omega_m)$ represents the normal-state self energy, which is given by
\begin{align}
\Sigma(\mathbf{p},i\omega_n) = -T_C \sum_{m\mathbf{k}} \frac{ V(\mathbf{p}-\mathbf{k},i\omega_n-i\omega_m)} {i\omega_m-\epsilon_\mathbf{k}-\Sigma(\mathbf{k},i\omega_m)}
\label{Eq_gap3}
\end{align}
In the above expressions the energy dispersion $\epsilon_\mathbf{k}$ is measured from the Fermi energy. Note that we will now restrict our attention to the conduction band intersecting the Fermi energy for electron doped system. We point out that one may include the phonon contribution in this framework by adding the phononic propagator $D(\mathbf{q},\omega)$ into the gap equations. Since, we are interested in evaluating the pure electronic contribution, we do not attempt this calculation here and reserve it for future studies.

\begin{figure}
    \centering
    \includegraphics[scale = 0.63]{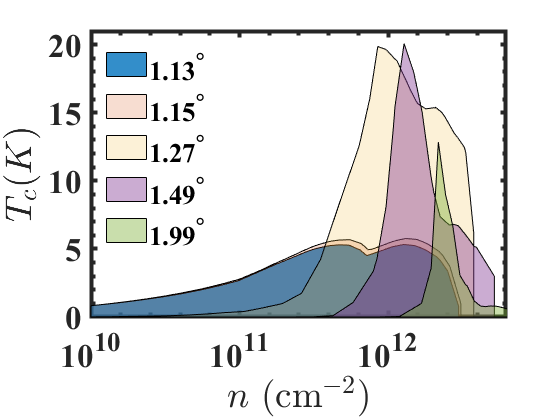}
    \caption{Critical temperature $T_C$ as a function of electron density in tBG (close to the magic angle). The presence of an asymmetrical superconducting dome around $n = 10^{12}$cm$^{-2}$ and $T_C = \mathcal{O}(10K)$ are the main predictions of our theory.}
    \label{fig:tc}
\end{figure}
\begin{figure}
    \centering
    \includegraphics[scale = 0.37]{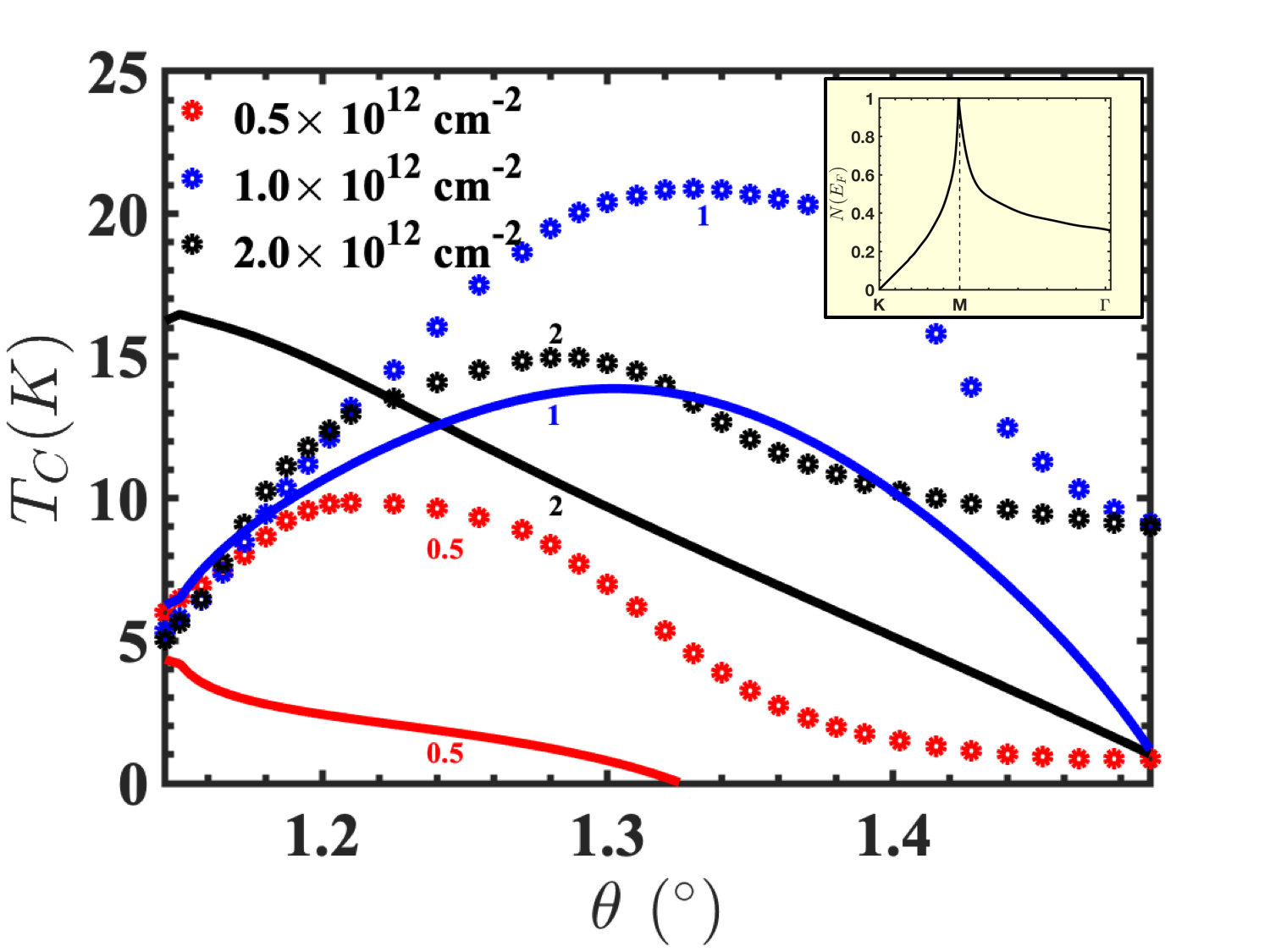}
    \caption{Critical temperature $T_C$ as a function of twist angle in tBG for three different carrier densities. The solid lines show the $T_C$ obtained from Eq.~\ref{TC} using the parameters $\omega_b$, $E_F$, and $\mu (r_s)$ from the numerical solution.
    The inset shows the normalized density of states (for any angle) at the Fermi energy as the Fermi surface moves up in energy and intersects various symmetry points in the Brillouin zone $\mathbf{K}\rightarrow\mathbf{M}\rightarrow \mathbf{\Gamma}$.}
    \label{fig:tc2}
\end{figure}

It has been shown~\cite{grabowski1984superconductivity} that superconductivity in a Coulomb system with no attractive interactions is essentially determined by the frequency dependence of the screened Coulomb interaction. Therefore, we can work with an effective Coulomb interaction, which is averaged over the momenta up to $\mathbf{k}_c = 2 \mathbf{k}_F$~\cite{grabowski1984superconductivity}, thus retaining only the crucial frequency dependence.  The momentum averaged interaction is given by 
\begin{align}
    \langle\langle V(i\omega_l)\rangle\rangle = \frac{\sum\limits_{\mathbf{k}, \mathbf{p}} \Theta(\mathbf{k}_c - \mathbf{k}) \Theta(\mathbf{k}_c - \mathbf{p}) V(\mathbf{p}-\mathbf{k}, i\omega_l)}{\sum\limits_{\mathbf{k}, \mathbf{p}}\Theta(\mathbf{k}_c - \mathbf{k}) \Theta(\mathbf{k}_c - \mathbf{p})}.
    \label{Eq:Veff_momentumavg}
\end{align}
The dimensionless coupling is given by $\lambda(i\omega)=\langle\langle V(i\omega)\rangle\rangle N(E_F)$, where $N(E_F)$ is the density of states. The limiting cases of the coupling are given by $\lim_{\omega = 0} \langle\langle V(i\omega)\rangle\rangle N(E_F)\rightarrow 0$ and $\lim_{\omega\rightarrow \infty} \langle\langle V(i\omega)\rangle\rangle N(E_F) = \mu$, where we define $\mu$ to be the high frequency limit of the coupling. To proceed further, the dimensionless coupling is then mapped onto the following explicit expression: 
\begin{align}
    \lambda_{nm}&=\langle\langle V(i\omega_n - i\omega_m)\rangle\rangle N(E_F) \nonumber\\
    &= \mu \left( 1 - \frac{\omega_b^2}{(\omega_n - \omega_m)^2 + \omega_b^2}\right),
    \label{Eq:lambda}
\end{align}
where $\omega_b$ is a boson  frequency that sets the scale of transition from the low to the high frequency limit. The mapping of the kernel onto the above Lorentzian form allows for an analytical treatment that is indeed found to resemble closely to the actual numerical solution for $\lambda_{nm}$. The parameters $\mu$ and $\omega_b$ are then extracted by fitting the actual coupling $\lambda(i\omega)$ to Eq.~(\ref{Eq:lambda}). The propagating boson here (with frequency $\omega_b$) is a collective density-fluctuation excitation such as electron-hole excitation or a plasmon. %We then consider~\cite{grabowski1984superconductivity} the following form of our interaction: $N(E_F) V(\mathbf{p}-\mathbf{k}, i\omega_n - i\omega_m) = \lambda_{nm} \Theta(k_c - |\mathbf{k}|) \Theta(k_c - |\mathbf{p}|) $
For the  momentum averaged interaction, Eq.~(\ref{Eq_gap1}) becomes momentum independent and is
given by Eq. (\ref{Eq:gap4}) with $Z_n = 1 + \mu (\omega_b/\omega_n) \arctan\left\{\omega_n E_F/ \left[(\omega_n^2 + \omega_b(E_F+\omega_b)\right]\right\}$ accounting for the self-energy corrections

The gap equation is now of the form of an eigenvalue equation $\bar{\Delta} = \hat{C}\bar{\Delta}$. At $T_C$ the largest eigenvalue of $\hat{C}$ is exactly one. Equation (\ref{Eq:gap4}) must be solved numerically to obtain a reliable value for critical temperature. The rate of convergence of the numerical solution depends on the ratio $T_C/E_F$. If $T_C\ll E_F$, the dimensions of the matrix involved can become prohibitively large to allow a numerical solution. However, in this regime the pseudopotential method~\cite{grabowski1984superconductivity}, which assumes $T_C\ll E_F$ as a premise, gives us a good estimate of the superconducting critical temperature. For our calculations we use a combination of numerical solution and pseudopotential method depending on the ratio $T_C/E_F$. For $\theta$ close to $\theta_M$, the numerical solution converges within a reasonable computational time. As $\theta$ increases the pseudopotential method is used. 
We also briefly comment on the nature of the superconducting gap function $\Delta (i\omega)$. Since we have purely repulsive interactions (unlike the case of phonon mediated BCS coupling), the gap function changes sign as a function of the Matsubara frequency. This is necessary for the gap equations to have a non-trivial solution. 

\textit{Superconductivity:} We will now discuss salient predictions of our theory. In Fig.~\ref{fig:tc} we plot the critical temperature $T_C$ as a function electron density in tBG close to the magic angle. The presence of an asymmetrical superconducting dome around $n = 10^{12}$ cm$^{-2}$ and the calculated $T_C = \mathcal{O}(10 K)$ are the main predictions of our theory. 
The obtained $T_C$ closely resembles the measured order of magnitude in experiments~\cite{cao_unconventional_2018, cao_correlated_2018,yankowitz_2019_tuning}, and also the calculated $T_C$ from the phonon contribution~\cite{lian2019twisted}. As the carrier density is increased the first maximum of the superconducting dome occurs when the Fermi energy intersects the $\mathbf{M}$ of the Brillouin zone point, where the van Hove singularity occurs. As the density is increased further, the dome is found to be asymmetric around this point, which can be understood from the fact that the density of states is not symmetric around the $\mathbf{M}$ point (see inset of Fig.~\ref{fig:tc2}). This asymmetry combined with effect of strong electronic interactions results in a second neighbouring maximum in the dome, which is more prominent at lower twist angles. When $\theta$ is increased further the width of the dome shrinks and the second maximum becomes less prominent since the electronic interactions become comparatively weaker. We observe superconductivity even close to $\theta\sim 2^{\circ}$, but the corresponding density window is quite narrow. In the monolayer limit the superconducting dome is practically non-existent within our model. The appearance of these domes is a special feature which arises from chosen realistic lattice model as opposed to the simple Dirac approximation for tBG. By examining the dependence on dielectric constant, we notice that the magnitude of $T_C$ is set primarily by $r_s$ (or $\mu$), while the shape of the domes is set by the density of states of the non-interacting bands.
Also note that we specifically focus on superconductivity. Other competing states (such as density waves, magnetism etc) may obviously affect the shape of the dome, when taken into consideration. 
In Fig.~\ref{fig:tc2} we plot the obtained critical temperature $T_C$ as a function of twist angle for different carrier densities. %The plateau near $\theta\sim 1.1^\circ$ is because of the weak bandwidth dependence near $\theta\sim 1.1^\circ$ as per the continuum model~\cite{bistritzer2011moire}. 
The dependence is observed to be non-monotonic, and for large angles $T_C$ is eventually suppressed, as expected. Fig.~\ref{fig:tc2} also compares the numerical solution to the $T_C$ obtained from single-mode model in Eq.~(\ref{TC}).

\textit{Concluding remarks:} We first point out that the vertex corrections are neglected in the Migdal-Eliashberg formalism, which may have a quantitative impact on the calculated $T_C$. 
For twist angles close to the magic angle, we find that the propagating boson frequency ($\omega_b$) can be several times larger than the Fermi energy $E_F$. However in this regime, the vertex corrections can be ignored, because they turn out to be insignificant for processes much larger than $E_F$ as pointed out earlier in the literature~\cite{takada1992insignificance}. Therefore the calculated $\mathcal{O}(T_C)$ is expected  to be robust especially close to the interesting regime of the magic angle. Vertex corrections may become more important for large twist angles causing suppression of $T_C$, but evaluating them is beyond the scope of the current manuscript.

To conclude, this work specifically points out to an important mechanism which is likely to be at play in superconducting tBG close to the magic angle. Collective excitations of strongly coupled electrons can mediate pairing, which may be dubbed as `plasmonic superconductivity', although the requirement of undamped plasmons is not strict in our formalism. We predict features of the theory, namely the superconducting dome and the magnitude of $T_C$, which are in good agreement with recent experimental data~\cite{cao_unconventional_2018, cao_unconventional_2018, yankowitz_2019_tuning}. The nature of the gap function in momentum space may be inferred by solving the Eliashberg equations without any momentum averaging. We reserve this technically harder problem for future studies. 

\textit{Acknowledgement:} G.S. acknowledges useful discussions with D. Y. H. Ho, N. Raghuvanshi, H-K. Tang, I. Yudhistira, N. Chakraborty, and J. N. Leaw.  This work was supported by the Singapore Ministry of Education AcRF Tier 2 grants MOE2017-T2-2-140 and MOE2017-T2-1-130.  M.T. acknowledges the Director's Senior Research Fellowship 
from the Centre for Advanced 2D Materials at the National University of Singapore 
(Singapore NRF grants R-723-000-001-281 and R-607-000-352-112).
O.P.S. and S.A. are supported by the Australian Research Council Centre of Excellence in Future Low-Energy Electronics Technologies (CE170100039).

\textit{Author contributions:} G.S. performed all the computations with inputs from M.T, G.V, and S.A.
 M.T. devised the single-mode model.  The project was conceived by G.S, O.P.S, G.V, and S.A.

\bibliographystyle{apsrev4-1}
\bibliography{biblio.bib}
\end{document}